\documentclass[%
preprint,onecolumn,
groupedaddress,
runinaddress,
frontmatterverbose,
showpacs,showkeys,
bibnotes,amsmath,amssymb,aps,
prl,
]{revtex4-1}

\usepackage[T2A]{fontenc}
\usepackage{amsfonts,amssymb,amsmath}
\usepackage{amsmath,graphicx}
\usepackage{graphicx,pifont}
\usepackage{citehack}
\usepackage{indentfirst}
\usepackage{commath}

\usepackage{graphicx}
\usepackage{dcolumn}
\usepackage{bm}
\usepackage{textcomp}
\usepackage[mathlines]{lineno}

\usepackage[usenames,dvipsnames]{pstricks}
\usepackage{epsfig}

\begin{document}
%
\title{Exactly solvable model of the ferroelectrics hysteresis loops control by the external electric field}
\author{A.Yu. Zakharov}\email[E-mail: ]{Anatoly.Zakharov@novsu.ru}
\author{M.I. Bichurin}\email[E-mail: ]{Mirza.Bichurin@novsu.ru}
\affiliation{Novgorod State University, Veliky Novgorod, 173003, Russia}

\author{Shashank Priya}\email[E-mail: ]{spriya@vt.edu}
\author{Yongke Yan}\email[E-mail: ] {yanthu@gmail.com}
\affiliation{CEHMS, Virginia Tech, Blacksburg, Virginia 24061, USA}
%
%
\begin{abstract}
The kinetic theory of switching processes in crystalline ferroelectric materials under the influence of a variable external electric field is formulated. The basic equations are derived and their exact analytical solution at arbitrary time-dependent external field are obtained. Connection between the hysteresis loops shape and variable external electric field parameters is investigated numerically. The numerical results were found to be in excellent agreement with the experiments.
\end{abstract}
%

\maketitle


\section{Introduction}
The origin  of the hysteresis phenomena in the condensed systems, including ferroelectrics, is existence of persistent metastable states. The relation between polarization of a ferroelectric~$ P\left( t\right)  $ and an external electric field strength~$E \left(t \right) $ has non-local character and depends not only on temperature and properties of a material, but also on the explicit view of function~$E(t)$, where $t$~is the time.

Control of the hysteresis shape of a ferroelectric material through an external electric field is possible if the connection between the shape of a variable external field and dynamics of domains switching processes can be established.

There are various approaches for analyzing the switching processes in ferroelectric materials. 
\begin{itemize}
\item The detailed description provided by the Ginzburg-Landau Hamiltonian can be based on the multidimensional Fokker-Planck equation~\cite{Kukushkin,Kaupuzs}. 
\item The finite element method~\cite{Dong} based on the phase field modelling with a Landau-Devonshire type multi-well potential. In this approach, the time-dependent Ginzburg-Landau equation with polarization as an order parameter governs the evolution of polarization and the domain wall width is controlled by a balance between mechanical, structure, electrostatic, and local gradient contributions to the free energy density. 
\item The inhomogeneous field mechanism (IFM) model~\cite{Tagantsev} developed primarily for polymer ferroelectrics~\cite{Genenko,Schutrumpf}. This model is based on the assumption that the switching volume can be divided into many spatial regions with {\em independent dynamics}, only determined by the local electric field. The local field values are random variables with distribution function related  to intrinsic inhomogeneities of the material.
\item There has been significant progress based on piezoresponse force microscopy used for imaging, manipulation and spectroscopy of ferroelectric switching processes. The review on this approach has been covered in Ref~\cite{Kalinin}. 
\item Modelling the hysteretic behavior of ferroelectric ceramics~\cite{Zak4} on the basis of the simplified model which neglects all the transitions from metastable states, except metastability boundaries. Results of modelling were in qualitative agreement with experimental data~\cite{Priya1,Priya2}. Through this simplified model  dependence of the hysteresis curves on external electric field frequency cannot be fully predicted.
\end{itemize}

This paper is devoted identification of the connection between~$P\left( t \right)$ and $E\left( t \right)$ for crystalline  ferroelectrics for a case of the arbitrary dependence of an electric field strength on time. 
The solution of this problem is fulfilled for the following simple model of switching processes in ferroelectric material:
\begin{enumerate}
\item It is supposed that the probability of switching of a single domain from metastable states into stable states depends on an external electric field strength and obeys to some arbitrary (but known) function $\alpha \left( E \right)$.
\item The switching processes of domains take place independently from each other.
\item Relation between the order parameter of a single domain and an external electric field strength is arbitrary (but known).
\end{enumerate}

The exact analytical solution of the model is presented. Using an illustrative example numerical solutions were obtained to validate the model. These numerical calculations are fulfilled for a case describing the connection between an order parameter of single domains and an external electric field in effective field model~\cite{ZBE}, and also for some model expressions for decay of metastable states probability~$\alpha$ as function of an external electric field strength~$E$.

\section{Kinetics of the switching  processes in the uni-axial ferroelectrics}

\subsection{The equations of a kinetics and their exact analytical solution}
\begin{figure}
\includegraphics[width=4.0in]{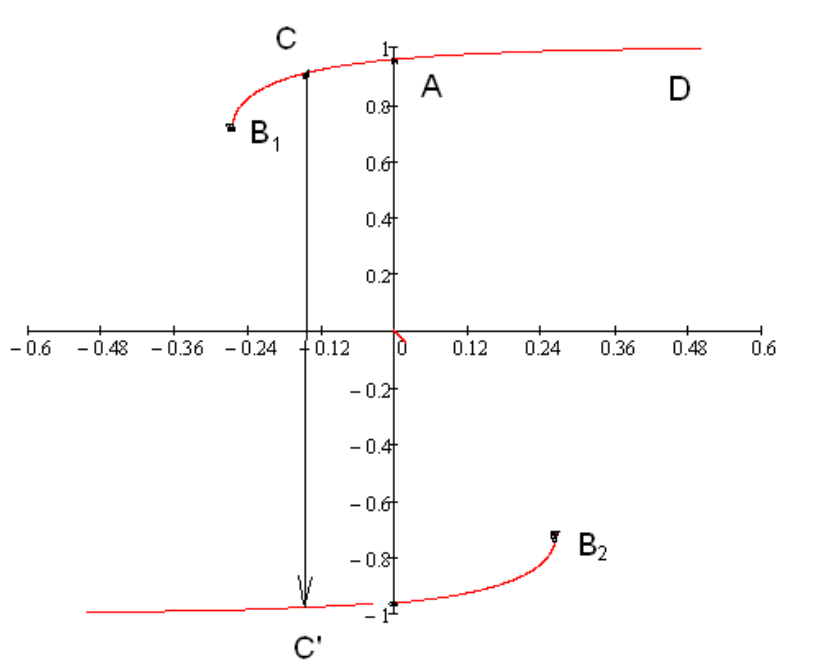}
\caption{Qualitative view of the dependence of the domain order parameter on an external electric field. }
\label{fig:1}
\end{figure}

Let's consider a model of the uni-axial crystalline ferroelectrics in which all~$N$ domains have a common direction of their axis. Each of the domains is characterized by an proper value of the order parameter (a local order parameter). The controlling external time-depending electric field~$ \mathbf {E} (t) $ is directed along the common axis of all domains. Those domains, for which at present time~$ t $ polarization~$ \mathbf {P} \left (\mathbf {E} \right)$ is parallel to external field, are in stable thermodynamic states. The domains with an opposite direction of polarization are in metastable states. We shall suppose, that both {\em univalent} functions~$P_{\pm} \left (E \right)$, defining connection between an external electric field and polarization of the domain, are known. The qualitative view of functions~$ P _ {\pm} \left (E \right) $ is shown in~Fig.\ref{fig:1}.

The electric field strength is placed on the abscissa axis. Point~$B$ denotes the metastability boundary. The line~$ \mathrm {AD}$ represents the stable equilibrium states of the domain. The line~$AB_1$ consists of metastable states: processes of switching of domains occur if the domains are placed in the states belonging to line~$AB_1$ of the curve. The probability of the domains switching per unit time is vanishing at the point~$A$ and tends to infinity at the metastability boundary point~$B_1$. The curve~$P_{-} \left (E \right)$ can be obtained from~$ P _ {+} \left (E \right) $ by the replacement $E \rightarrow-E$ and $P \rightarrow -P $.

\begin{figure}
\includegraphics[width=4.0in]{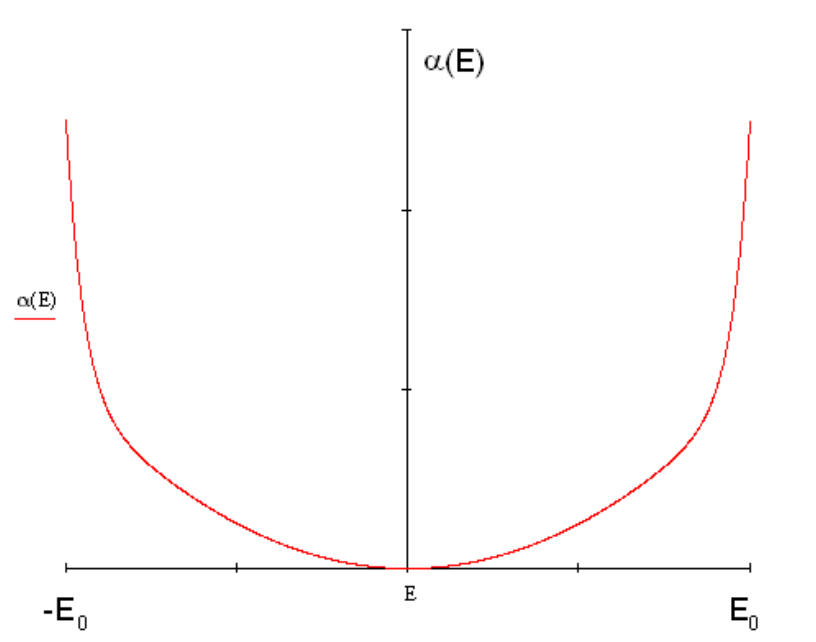}
\caption{Qualitative view of the dependence of rate of decay of metastable states as a function of an external field $E$.}
\label{fig:2}
\end{figure}

Probability of the domain switching per unit time from an external field is some even function~$\alpha \left (E \right)$ which also depends on temperature and ferroelectric characteristics. The qualitative graph of this function is shown in Fig.\ref {fig:2}.

Let's choose one of the possible directions of polarization of dipoles as the positive direction. We will denote the boundary of metastability by~$E_0$. Then at~ $-E_0 <E \left (t \right) <E_0 $ the part of domains belong in the metastable states and pass into the stable states.

Assume, that the external field changes in the course of time under the known function
\begin{equation}\label{E(t)}
{E} = {E}\left( t\right).
\end{equation}
Then evolution of the domains in time obeys to following system of the ordinary differential equations
\begin{equation}\label{evolution}
\left\{
\begin{array}{l}
{\displaystyle  \frac{dN_1}{dt} = \alpha\left( {E} \right)\left[ -\theta\left(- {E} \right) N_1 + \theta \left( {E} \right) N_2 \right];   }\\  \\
{\displaystyle  \frac{dN_2}{dt} = \alpha\left( {E} \right)\left[ \theta\left(- {E} \right) N_1 - \theta \left(  {E}\right) N_2 \right], }
\end{array}
\right.
\end{equation}
where~$N_1\left(t \right)$ and $N_2\left(t \right)$ are numbers of positively and negatively oriented domains, respectively,
 $\theta\left( x\right)$ is the Heaviside step-function given as:
\begin{equation}\label{heavy}
\theta \left( x \right) =   \left\{
\begin{array}{l}
{\displaystyle  1, \qquad x > 0;  }\\
{\displaystyle 0.5,\, \quad x = 0;   }\\
{\displaystyle 0, \qquad x < 0.   }
\end{array}
\right.
\end{equation}
It should be noted that
\begin{equation}\label{N1+N2}
N_1\left(t \right) + N_2\left(t \right) = \mathrm{const} = N.
\end{equation}

Excluding in the system of equations~(\ref {evolution}) the function $N_2 \left (t \right) $, we obtain the following  equation with respect to fraction of positively oriented domains~$n_1\left(t \right) = \frac{N_1\left(t \right)}{N}$:
\begin{equation}\label{dn1}
    \frac{dn_1\left( t\right)}{dt} + \alpha\left( {E} \right)\, n_1\left( t\right) = \alpha\left( {E} \right)\, \theta \left( {E} \right).
\end{equation}

The Cauchy problem for differential equation~(\ref{dn1}) with condition 
\begin{equation}\label{n10}
    \left. n_1\left(t \right) \right|_{t=0} = n_1\left(0 \right)
\end{equation}
has the following solution
\begin{equation}\label{n1t}
\begin{array}{r}
    {\displaystyle     n_1\left(t \right) = \left[\int\limits_0^t \theta \left ( {E} \left(t_2 \right) \right ) \alpha \left ( {E}\left(t_2 \right) \right )\,  \exp \left ( \int\limits_0^{t_2} \alpha\left( {E} \left(t_1 \right) \right) dt_1 \right ) dt_2 + n_1\left(0 \right) \right]  }\\
{\displaystyle \times   \exp \left (- \int\limits_0^t  \alpha\left( {E} \left(t_3 \right) \right)\, dt_3  \right ).  }
\end{array}
\end{equation}

Let's introduce the notation
\begin{equation}\label{F(t)} 
    F\left( t \right) = \int\limits_{0}^{t} \alpha\left( {E}\left( t_1\right) \right)\, d t_1.
\end{equation}
$F\left(t \right)$ is nondecreasing function as far as
\begin{equation}\label{F'(t)}
   \frac{dF(t)}{dt} = \alpha\left( {E}\left( t \right) \right) \ge 0,
\end{equation}
At sufficiently large values of~$t$, function~$\exp\left( - F\left(t \right)\right)$  tends to zero, therefore the system ``forgets'' their initial state~$n_1 \left (0 \right)$. Thus, at~$ t \gg f ^ {-1} $, where $f $ is frequency of an external field, the solution~(\ref {n1t}) achieves the following form
\begin{equation}\label{n1(t)}
     n_1\left(t \right) = \exp \left (- F \left(t \right)  \right )\, \left[\int\limits_0^t \theta \left ( {E} \left(t_1 \right) \right ) \alpha \left ( {E}\left(t_1 \right) \right )\,  \exp \left ( F\left(t_1 \right) \right ) dt_1 \right].
\end{equation}
It is easy to see that $n_2\left(t \right)$ can be expressed via $n_1\left(t \right)$:
\begin{equation}\label{n2(t)}
    n_2\left(t \right) = 1 - n_1\left(t \right).
\end{equation}

\subsection{Analysis of the kinetic equation solution for sinusoidal external field}

Formally exact solution given by Eq.~(\ref{n1(t)}, \ref{n2(t)}) through the evolution equations~(\ref {evolution}) in general form does not permit a detailed qualitative analysis.  The qualitative (and all the more~ --- the quantitative!) behaviour of solution~(\ref{n1(t)}) depends essentially on an explicit form of function~$ \alpha \left (E \right) $.

One of the variants of analytical representation of function~$ \alpha \left (E \right)$ consists of the following. We will rewrite function~$ \alpha \left ({E} \right)$  in the form of the sum of two terms, first of which dominates in vicinity of the point~$E=0$ (it is approximated by quadratic functions~ $ E / {E_0} $), and the second term dominates in vicinity of the points~$E = \pm E_1^{(0)}$ (it is approximated by a high degree from~$ E / {E_0} $):
\begin{equation}\label{alphaE}
    \alpha\left( {E}\right) = a_1 \left(\frac{{E}}{E_1^{(0)}} \right)^2 + a_2 \left(\frac{{E}}{E_1^{(0)}} \right)^{2m},
\end{equation}
where $a_1$ and $a_2$ are some constants.

Thus, we shall choose function~$ \alpha \left (E \right) $ in the form of expression~(\ref {alphaE}), containing three parameters $a_1, \ a_2, \ m $. Greater  magnitude of parameter~$m$ is provided with the complete devastation of metastable states on achievement of an external field of values~$\pm E_0 $.

\begin{figure}
\includegraphics[width=4.5in]{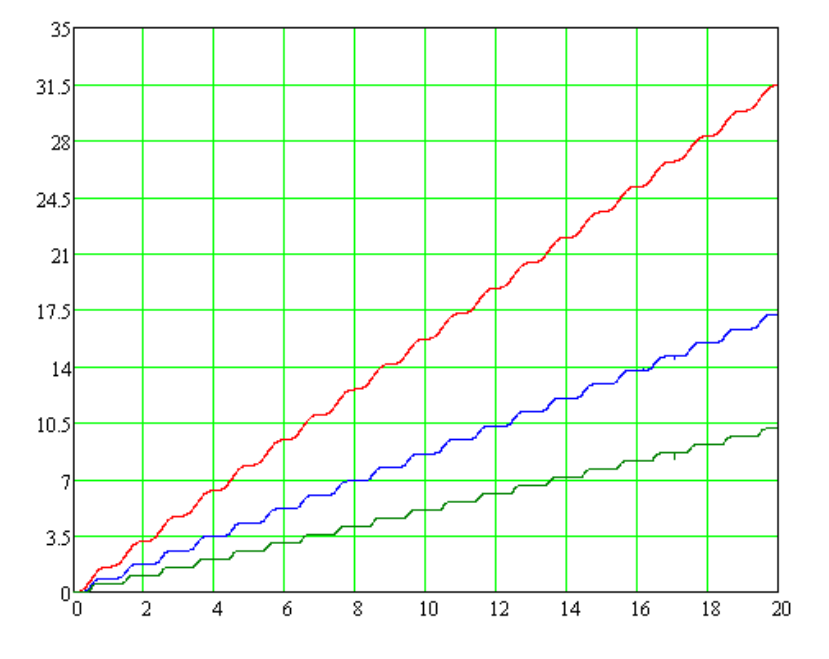}
\caption{Qualitative view of the dependence of functions~$F\left (n, t \right) $ (ordinate axis) on time~$ t $ (abscissa axis) at $n=1, \, 4, \, 12$ (red, blue, and green lines, respectively).}
\label{fig:3}
\end{figure}

\begin{figure}
\includegraphics[width=4.5in]{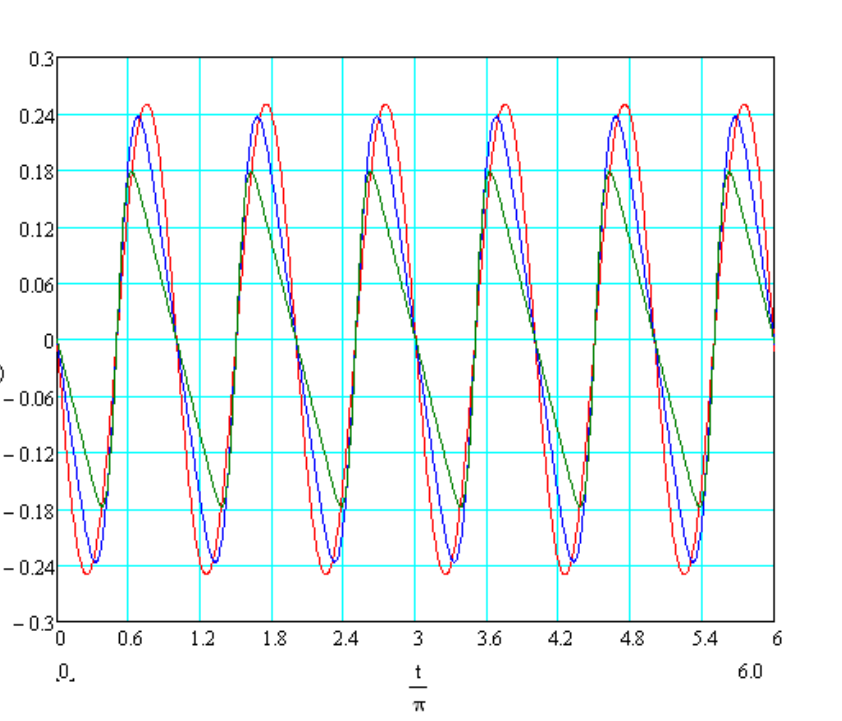}
\caption{Qualitative view of an oscillating components of functions~$ F\left (n, t \right)$ (ordinate axis) on time~$ t $ (abscissa axis)  at $n=1, \, 4, \, 12$ (red, blue, and green lines, respectively).}
\label{fig:4}
\end{figure}

Substitution of expression~(\ref{alphaE}) into~(\ref{F(t)}) for an external field changing under the harmonic function
\begin{equation}\label{sinot}
{E}\left( t\right) = A\, \sin\left(\omega t \right),
\end{equation}
leads to integrals which are calculated in elementary functions. The result of integration contains the linear $ {\displaystyle \frac {\left (2n \right)!} {\left [2^n \, n! \right] ^2} \, t} $ and oscillating functions of time:
\begin{equation}\label{sin2n}
  F\left(n, t \right) = \int\limits_0^t \sin^{2n}(\omega t)\, dt = \frac{\left( 2n \right)!}{\left[2^n\, n! \right]^2}\, t + \frac{(-1)^n}{2^{2n}\,\omega}\, \sum_{k=0}^{n-1}\, (-1)^k\, C_{2n}^{\, k}\, \frac{\sin\left[2(n-k)\omega t\right]}{n-k},
\end{equation}
where $C_{2n}^{\,k} = \frac{(2n)!}{k!\,(2n-k)!}$ are the binomial coefficients.
The typical graphs of functions~$F \left (n, t \right)$ are presented in Fig.\ref {fig:3}.
The graphs of oscillating parts of functions~$ F \left (n, t \right) $ at $n=1, \, 4, \, 12 $ are presented in Fig.\ref {fig:4}.

Thus, we have the following expression for function~$ F \left (t \right) $:
\begin{equation}\label{F1(t)}
\begin{array}{r}
    {\displaystyle  F\left(t \right) = \int\limits_{0}^{t} \alpha\left(  {E}\left(t' \right)\right)\, dt' = a_1\, \left(  \frac{A}{E_0} \right)^2 \left[\frac 1 2\, t + \varphi_2\left(t \right) \right]  }\\
{\displaystyle    + a_2\, \left( \frac{A}{E_0} \right)^{2m} \left[ \frac{(2m)!}{\left[2^m\, m! \right]^2}\, t + \varphi_{2m}\left(t \right) \right]  ,  }
\end{array}
\end{equation}
where $\varphi_{2n}\left(t \right)$~ are the periodic functions of time defined by the relation:
\begin{equation}\label{phint}
    \varphi_{2n}\left(t \right) = \frac{(-1)^n}{2^{2n}\,\omega}\, \sum_{k=0}^{n-1}\, (-1)^k\, C_{2n}^{\, k}\, \frac{\sin\left[2(n-k)\omega t\right]}{n-k}.
\end{equation}

At large values of time $t $ the expression~(\ref {n1(t)}) contains the product of infinitesimal function $ \exp \left (-F \left (t \right) \right) $ and infinite function $\exp \left (F \left (t_1 \right) \right) $ in the integrand. Therefore for numerical calculations it is appropriate to use the following representation for~$ n_1(t) $: 
\begin{equation}\label{n1tdn1t}
    n_1\left( t\right) = \int\limits_0^t dt_1\ \theta \left ( {E} \left(t_1 \right) \right )\, \alpha \left ( {E}\left(t_1 \right) \right )\, \exp \left [ F\left( t_1 \right) - F\left( t \right) \right ].
\end{equation}

\begin{figure}
\includegraphics[width=4.5in]{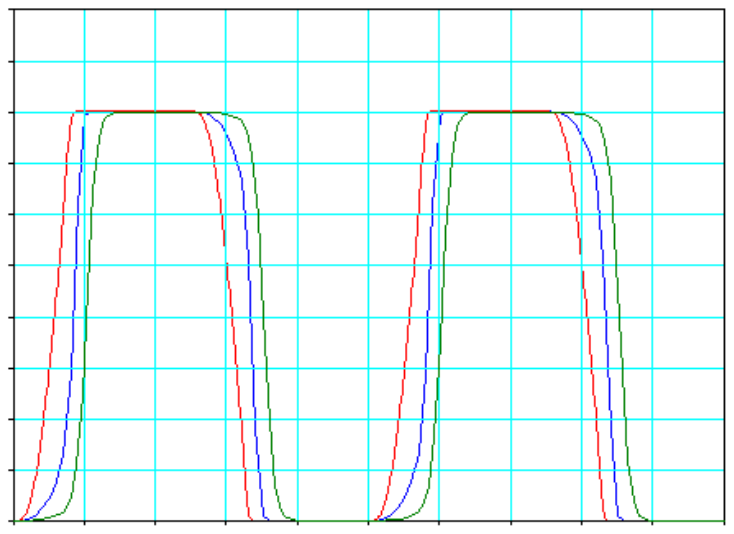}
\caption{Dependence of the positively oriented domains fractions~$ n_1$ (ordinate axis) on quantity~$ f t $ (abscissa axis) at the dimensionless frequencies of a controlling field $f=0.25, \ 2.0, \ 16$, (red, blue, and green lines, respectively).}
\label{fig:5}
\end{figure}

By way of illustration, we present graphs of dependence~$n_1 $ from product~$ \frac {\omega} {2 \pi} t = f \cdot t $ for frequencies~$ f=0.25, \, 1.0, \, 3.0, \, 7.0 $ at a variable external field~(\ref {sinot}). Here we choose the following values of the dimensionless parameters~$ \frac {A} {E_0} =1.4 $~(the amplitude of an external field should be more, than the boundary of metastability~$ A> \left|E_0 \right|  $), $a_1=5 $, $a_2=1.5 $, $m=10 $ (in an ideal way~$ m \to \infty $, in a reality $m \gg 1 $). Results of numerical calculations are presented in Fig.\ref {fig:5}.

As we can see from these calculations, the change of the shape of a curves~$n_1(t) $ are caused not only by change of a time scale: in variables~$ f\cdot t-n_1 $ the shape of a curve undergoes appreciable changes with  increasing frequency. The physical reason for this phenomenon is following: at low frequencies the system transits slowly from the range of small strength of the external field and during this time even infrequent switchings of domains have enough time for accumulation. With increasing frequency  most of the switchings relate to the external fields values in a neighbourhood of metastability boundary~$\pm E_0$. This regularity becomes especially obvious if we were to carry out these calculations in variables~$n_1 (t) $~---$E (t) $.

\begin{figure}
\includegraphics[width=4.5in]{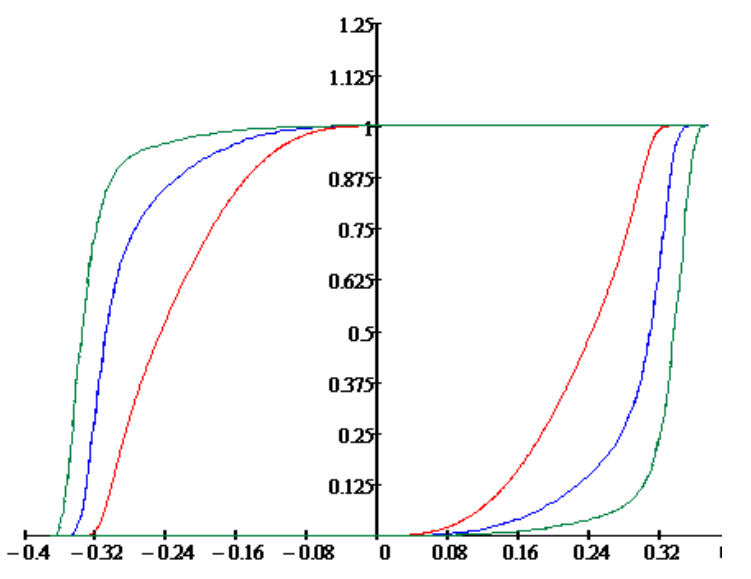}
\caption{Hysteresis curves of positively oriented domains fraction $n_1$ on electric field strength $E$ at the dimensionless frequencies of a controlling field $f=0.25, \ 1.0,\ 4 $ (red, blue and green lines, respectively). }
\label{fig:6}
\end{figure}

Evolution of the shape of the hysteresis curves in form of the  variables ``fraction of positively oriented domains''~---`` external field strength'' with change of frequency is shown in Fig.\ref {fig:6}.

Note that with increasing controlling field frequency there is an change of a hysteresis curves~$ n_1-E $ to the rectangular shape. However, our basic interest are the hysteresis curves for polarization.


\section{hysteresis curves of ferroelectrics for sinusoidal external field}

Relation between functions~$n_1(t)$ and~$E(t)$ has the following form
\begin{equation}\label{P(t)E(t)}
    P\left(t \right) = n_1\left( t\right)P_{+}\left(E(t) \right) + \left[1 - n_1\left( t\right) \right] P_{-}\left( E(t) \right),
\end{equation}
where~$P_{\pm}\left(E \right)$~ are univalent functions, defining the dependence of an order parameter of the domain from a controlling field while taking account of metastable states. 

In general case, these functions are unknown and depend on the choice of a ferroelectric model. In the case of model with long-range interactions between the dipoles, interconnection between an external field and an order parameter of the domain is set by the relation~\cite{ZBE}:
\begin{equation}\label{E-p}
    \mathcal{E} = \frac{\tau}{2} \ln \left[\frac{1 + P}{1 - P} \right] - P
\end{equation}
where $\tau,\ P,\ \mathcal{E}$ are dimensionless temperature, order parameter, and external field strength, respectively:
\begin{equation}\label{defs}
    \tau = \frac{T}{T_c},\quad P = \frac{\left< p\right>}{p_0},\quad \mathcal{E} = \frac{Ep_0}{T_c},
\end{equation}
$T$ is the sample temperature, $T_c$~is the Curie temperature, , $\left< p \right>$~is mean value of the dipole moment of a cell, $p_0 $~is the dipole moment of unit cell magnitude.

The metastability boundaries are points~$ M ^ {(0)} _ {1} \left (p ^ {(0)} _ {1}, \, \mathcal {E} ^ {(0)} _ {1} \right) $ and $M ^ {(0)} _ {2} \left (p ^ {(0)} _ {2}, \, \mathcal {E} ^ {(0)} _ {2} \right) $ on a plane of variables~$ \left (P, \, \mathcal {E} \right) $ which are defined from a condition of vanishing of a derivative with respect to~$ P $ on the right hand side of~(\ref {E-p}):
\begin{equation}\label{metastab}
    \left\{
\begin{array}{l}
    {\displaystyle  p^{(0)}_{1,2} = \pm \sqrt{1 - \tau};  }    \\
{\displaystyle \mathcal{E}^{(0)}_{1,2}  = \frac{\tau}{2} \ln \left[\frac{1 + p^{(0)}_{1,2}}{1 - p^{(0)}_{1,2}} \right] - p^{(0)}_{1,2}\,.  }
\end{array}
\right.
\end{equation} 
Thus, in the given model the dimensionless values of both an external field and an order parameter on metastability boundary are uniquely determined by the value of the dimensionless temperature of the sample.

Functions~$ P_{\pm} \left (\mathcal {E} \right) $~ are two branches of the solution of the equation~(\ref{E-p}) with respect to~$P$ with ``link'' to points $M ^ {(0)} _ {1,2}$, respectively.  Unfortunately, exact analytical solution of this equation  is impossible, therefore we use the following method for determining the solution.

At first we reduce the equation~(\ref{E-p}) to the equivalent form and use ``link'' to point~$ M _ {1} ^ {(0)} $. As a result, we have the system of equations given as:
\begin{equation}\label{EE0}
\left\{
\begin{array}{l}
{\displaystyle  P_{+}\left(\mathcal{E} \right) = \tanh\left(\frac{\mathcal{E}+P_{+}\left(\mathcal{E} \right)}{\tau} \right);  }\\
{\displaystyle p_1^{(0)} \left(\mathcal{E}_1^{(0)} \right) = \tanh\left(\frac{\mathcal{E}_1^{(0)}+p_1^{(0)}\left(\mathcal{E}_1^{(0)} \right)}{\tau} \right). }
\end{array}
\right.
\end{equation}
Let us introduce the notation 
\begin{equation}\label{delta}
    \delta\left( \mathcal{E}\right) =  P_{+}\left(\mathcal{E} \right) - p_1^{(0)} \left(\mathcal{E}_1^{(0)} \right)
\end{equation}
and obtain the equation with respect to~$\delta\left( \mathcal{E}\right)$:
\begin{equation}\label{dE}
     \delta\left( \mathcal{E}\right) = \frac { \tau\, \tanh\left (\frac{\mathcal{E} - \mathcal{E}_1^{(0)} + \delta\left( \mathcal{E}\right) }{\tau} \right )} { 1 + p_1^{(0)}  \tanh\left (\frac{\mathcal{E} - \mathcal{E}_1^{(0)} + \delta\left( \mathcal{E}\right) }{\tau} \right ) }.
\end{equation}

As long the absolute value of a derivative of the right hand side of this equation with respect to~$ \delta \left (\mathcal {E} \right) $ is less than unity at all values of input variables and parameters, this equation can be solved by a simple iterations  method with any preassigned precision:
\begin{equation}\label{d(n+1)}
    \delta_{n+1}\left( \mathcal{E}\right) = \frac { \tau\, \tanh\left (\frac{\mathcal{E} - \mathcal{E}_1^{(0)} + \delta_n\left( \mathcal{E}\right) }{\tau} \right )} { 1 + p_1^{(0)}  \tanh\left (\frac{\mathcal{E} - \mathcal{E}_1^{(0)} + \delta_n\left( \mathcal{E}\right) }{\tau} \right ) }.
\end{equation}
Let the initial value of $\delta_{0}\left( \mathcal{E}\right) = 0$.
After $m$ iterations we obtain
\begin{equation}\label{p1(E)}
    P_{+}\left( \mathcal{E}\right) \approx p_1^{(0)} + \delta_m\left( \mathcal{E}\right).
\end{equation} 

Function~$P_{-}\left( \mathcal{E}\right)$ can be find using~$P_{+}\left( \mathcal{E}\right)$:
\begin{equation}\label{p2(E)}
    P_{-}\left( \mathcal{E}\right) = - P_{+}\left( -\mathcal{E}\right).
\end{equation}
Thus, we have all the necessary information to calculate the hysteresis curves of ferroelectrics.

\begin{figure}
\includegraphics[width=4.5in]{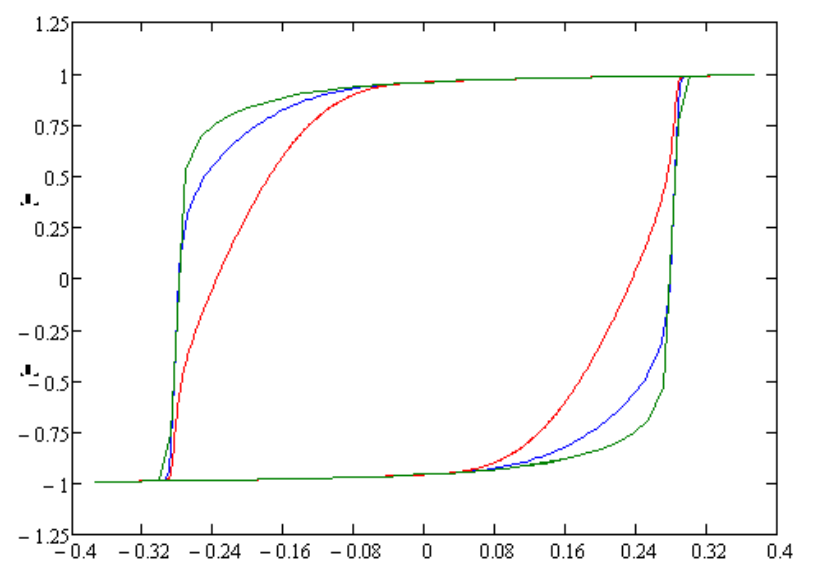}
\caption{Hysteresis curves of dimensionless polarization $P$ on dimensionless electric field strength $E$ at the controlling external field  frequencies $f=0.25, \ 1.0,\ 4.0 $ (red, blue and green lines, respectively).}
\label{fig:7}
\end{figure}

Using expressions~(\ref{p1(E)}) and (\ref{p2(E)}) for~$P_{\pm}\left(E(f,t) \right)$ and~(\ref{n1tdn1t}) for~$n_1\left(t \right)$ in the relation~(\ref{P(t)E(t)}) gives numerical evaluation of polarization. Here $\tau = 0.5$, $A_1=1.4$, $a_1 = 5.0$, $a_2 = 1.5$. Frequencies of the external field were chosen as $f = 0.25, \ 1.0,\ 3.0,\ 7.0$. Results of the numerical calculations are presented in Fig.\ref {fig:7}.

\section{Conclusion}

The results in this paper can be summarized as:
\begin{enumerate}
\item The kinetic equations of domains switching processes of the uni-axial crystalline ferroelectric material exposed to time-depending external electric field were derived.
\item Exact analytical solutions of the kinetic equations for arbitrary dependence of metastable states decay probability on the external electric field~$\alpha\left( E\right) $ and on the arbitrary dependence of an external field on time~$E\left( t\right) $ were obtained.
\item Numerical calculations for the  dependence of domains distribution function over their orientations as a function of time were obtained. It was shown that the hysteresis curves shapes depends essentially on the frequency of an external controlling field.
\end{enumerate}

Especially it should be noted that the shape of the hysteresis curves of a ferroelectric material is composite of two functions
\begin{enumerate}
\item variation of decay of metastable states probability as a function of an external electric field strength~$\alpha\left( E\right)$;
\item variation of external electric field strength as a function of time~$E\left( t \right) $.
\end{enumerate}

Thus, there are two types of the problems concerning to the modeling of the hysteresis phenomena in  ferroelectric materials.
\begin{enumerate}
\item the direct problem: control of the hysteresis curves shapes by means of external electric field~$E\left( t \right) $; it is possible if the function $\alpha\left( E\right)$ is known and this problem is interesting from the application point of view.
\item the inverse problem: Establishment of a functional dependence~$\alpha\left( E\right)$ from shapes of the hysteresis curves at different time dependent functions~$E\left( t \right) $; this approach requires information related to the microscopic mechanisms of switching processes in  ferroelectric materials.
\end{enumerate}

\newpage

\end{document}